\documentclass[preprint2]{aastex}

\newcommand{\myemail}{Martin.Still@gsfc.nasa.gov}
\newcommand{\her}{\mbox{Her X-1}}
\newcommand{\xte}{\mbox{\it RXTE}}
\newcommand{\cgro}{\mbox{\it CGRO}}
\newcommand{\dex}[1]{\hbox{$\times\hbox{10}^{#1}$}}
\newcommand{\kev}{\,\mbox{keV}}

\newcommand{\fmax}{\mbox{$F_{\mathrm {max}}$}}
\newcommand{\eprec}{\mbox{$E_{\mathrm {35}}$}}
\newcommand{\pprec}{\mbox{$P_{\mathrm {35}}$}}
\newcommand{\pprecdot}{\mbox{$\dot{P}_{\mathrm 35}$}}

\newcommand{\phip}{\mbox{$\phi_{\mathrm {35}}$}}
\newcommand{\phio}{\mbox{$\phi_{\mathrm {orb}}$}}
\newcommand{\cs}{\mbox{count s$^{-1}$}}

\slugcomment{Submitted to Astrophysical Journal Letters, Feb 2004}

\shorttitle{Recalibrating the clock in Her X-1}
\shortauthors{Still and Boyd}

\begin{document}

\title{Fine-tuning the accretion disk clock in Hercules X-1}

\author{M. Still\altaffilmark{1} and P. Boyd\altaffilmark{2}}
\affil{NASA/Goddard Space Flight Center, Greenbelt, MD 20771}
\email{\myemail, padi@milkyway.gsfc.nasa.gov}

\altaffiltext{1}{Universities Space Research Association.}
\altaffiltext{2}{University of Maryland Baltimore County.}

\begin{abstract}

{\it RXTE} ASM count rates from the X-ray pulsar Her X-1 began falling
consistently during the late months of 2003.  The source is undergoing
another state transition  similar   to  the  anomalous low state    of
1999. This new event  has triggered observations  from both space- and
ground-based observatories.  In order  to aid data interpretation  and
telescope scheduling, and to facilitate the phase-connection of cycles
before  and  after the  state   transition, we have  re-calculated the
precession ephemeris using cycles over  the last 3.5 years.  We report
that the source has displayed a different  precession period since the
last anomalous event.  Additional archival data from
\cgro\  suggests that each  low  state is accompanied  by a  change in
precession period and that   the subsequent period is correlated  with
accretion flux.  Consequently our analysis reveals long-term accretion
disk  behaviour  which    is predicted   by     theoretical models  of
radiation-driven warping.
\end{abstract}


\section{Introduction}
\label{sec:introduction}

Her X-1 is an eclipsing X-ray pulsar that displays variability on spin
(1.24s),   orbital  (1.7d)   and    super-orbital   (35d)   timescales
\citep{tan72}.  The companion star is   of early-F or late-A type  and
fills its  Roche lobe, resulting  in  accretion mainly  by Roche  lobe
overflow through the inner Lagrangian ($L_1$) point \citep{lea98}.

The 35d  period  is  revealed in  X-rays   as an absorbing   column of
variable depth passing in front of the accretion source \citep{sco00},
In the UV and optical, an intrinsic flux variation  is observed on the
same  period \citep{ger76,lea99}.   The  long-standing  model is  of a
warped accretion  disk,  viewed close to  edge-on ($i  \simeq 90$ deg)
precessing  in   a  retrograde  direction  around  the   neutron star,
\citep{kat73,rob74}.  The disk  atmosphere   provides the  column  and
casts an  X-ray  shadow over the  inner  face  of the  companion star,
causing optical variability on 35d timescales.

\begin{figure*}
\begin{picture}(100,0)(10,20) 
\put(0,0){\includegraphics{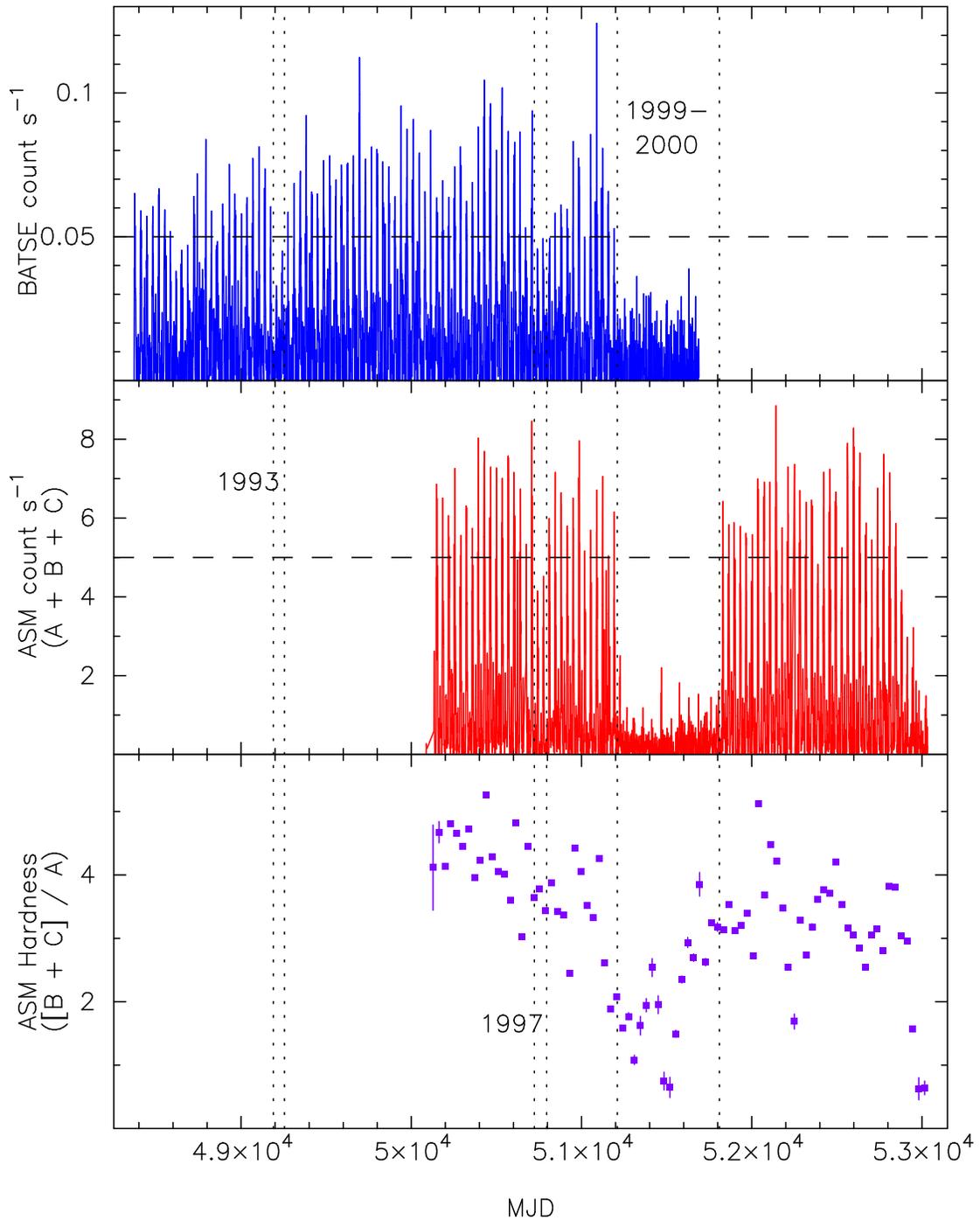}}
\noindent
\end{picture}
\vspace{190mm}
\figcaption{Top panel: \her\ \cgro\ BATSE (20--50 \kev) one-day average 
count  rates from Earth occultations  between 1991--1999. Middle panel:
\xte\  ASM (1--12  \kev) one-day average  count rates  from \her\ from
1996--2004.   Bottom panel: ASM Hardness  ratio  averaged over each 35
day cycle.  Band  A = 1.3--3 \kev,  band B =   3--5\kev\ and band C  =
5--12\kev.  Horizontal  lines represent count rate  thresholds adopted
in the $0-C$ analysis. Dwells  over the two  thresholds are plotted on
the $O-C$ diagram of Fig. \ref{fig:o-c}\label{fig:asm}}
\end{figure*}

In X-rays,  the 35d cycle   is  characterized by  a  rapid turn-on,  a
``main-on''  state  which varies  in intensity greatly   from cycle to
cycle, and then  decays back to an ``off''  state at $\sim 1$\% of the
peak main-on flux. This is followed by a rise to a weaker ``short-on''
state and then another low  \citep{tan72,gor82}. Superimposed on  this
cycle are eclipses of the neutron star by the companion once per orbit
and dips  that  occur  almost on the  beat   period between orbit  and
precession \citep{gia73}, which  are believed to  be due to  localized
structure in the  disk, generated by the impact  of the  ballistic gas
stream  from the $L_1$  point \citep{cro80,sch96}.  The turn-on occurs
at only two orbital phases, $\phio  \simeq 0.2$ or $\phio \simeq 0.7$,
apparently  at random.   Consequently  it has  often been claimed that
precession is not strictly   periodic (Ogelman 1987),   and precession
cycles generally  have durations  of either 20,   20.5  or 21  orbital
cycles,  leading  some authors  to  suggest  that  the precession  and
orbital cycles are related physically \citep{sco99}.

To date, it has not been possible  to determine the engine behind disk
precession   observationally.  From   a  theoretical  perspective, the
companion star can force a tilted disk to precess in the outer regions
closest to  the donor star \citep{lar98}. However  tidal forces do not
account for the  mechanism that tilts  the disk.   Mechanisms that can
simultaneously    warp   structure  and   drive   precession  are wind
\citep{sch94} and radiation \citep{pri96,wij99} pressure, pushing down
on  the  disk  plane.  The dominant  engine  behind  this pressure  is
accretion   flux from  the neutron  star    which irradiates the  disk
atmosphere  externally.  While  the  simplifying  assumptions used  in
these theoretical  calculations   prevent a detailed  comparison  with
observation, the success of the models  in producing warped disks with
the observed precession rate indicates that some combination of tides,
wind and radiation pressure are capable of driving the \her\ clock.

Her  X-1   has occasionally  displayed  consecutive main-on  states of
significantly  lower flux  than   average,  e.g.   1993 \citep{vrt94}.
Also,   between  1999--2000,  the source  missed   a  large number  of
consecutive cycles  altogether  \citep{par99,cob00,vrt01,sti01}.   See
also \citet{par85}  for the description  of a 1983  event prior to the
epoch of all-sky monitoring surveys.  During  these times, the optical
flux  does   not dip appreciably  \citep{vrt01}.  In   addition to the
decrease in main-on amplitude,  the neutron star experiences an abrupt
transition from the nominal state  of pulse period spin-up to episodes
of spin-down \citep{par99}.  Although these events are not homogeneous
in  terms of  either duration   or  magnitude, they have  been  coined
collectively the `Anomalous Low  States (ALS)'.  Quite clearly the ALS
are significant disk events; they point to the accretion disk changing
state  for a limited period of  time before returning seemingly to the
same preferred period and warp shape, albeit with an offset of several
orbital cycles in the   turn-on clock, according to  \citet{oos01} and
\citet{man03}. In this paper, we show this picture to be incorrect.

{\it Rossi X-ray Timing Explorer  (RXTE)} All-Sky Monitor (ASM)  count
rates from  Her X-1  in the  main-on state  began falling consistently
during  the late  months of  2003.   \her\  appears  to be  undergoing
another state  transition similar to the ALS  of 1999--2000.   Since a
glitch   in  the precession clock  was   reported  after the  last ALS
\citep{oos01}, it is now    prudent  to recalculate  the    precession
ephemeris using the last 3.5 years of  data.  Rather than a glitch, we
report  that the source  has displayed a  different precession period,
\pprec, since  the  last ALS.   Additionally, archival   data from the
BATSE  experiment that  was   onboard  the  {\it  Compton  Gamma   Ray
Observatory (CGRO)} suggests that each ALS  is accompanied by a change
in \pprec.  Contrary to  previous reports \citep{oos01}, these results
indicate  that the accretion  disk does not  return to  the same state
after each ALS.

\section{Results}
\label{sec:results}

Fig.~\ref{fig:asm}  plots   the total BATSE  and   ASM  count  rate as
averages  of  all  dwells taken  within   a  24h period,  weighted  by
uncertainty.  Three  ASM energy channels  allow  the source's spectral
evolution to  be monitored coarsely as  a hardness ratio. The main-ons
of the 35d  cycle reveal themselves  as maxima in  the count rate time
series, where each cycle peaks at 3--9 ASM count s$^{-1}$ (1 Crab = 75
\cs).  The stand-out  features of  this  time series are the  extended
1999 ALS between  MJD 51\,195--51\,892 \citep{par99}, and  the decline
into a putative new ALS starting  MJD 52\,866 \citep{boy04}.  The 1993
ALS reported by \citet{vrt94} can be  seen as two consecutive main-ons
with faint maxima of $< 0.05$ BATSE \cs\ between MJD 49\,213--49\,282.

\begin{figure*}
\begin{picture}(100,0)(10,20) 
\put(0,0){\includegraphics{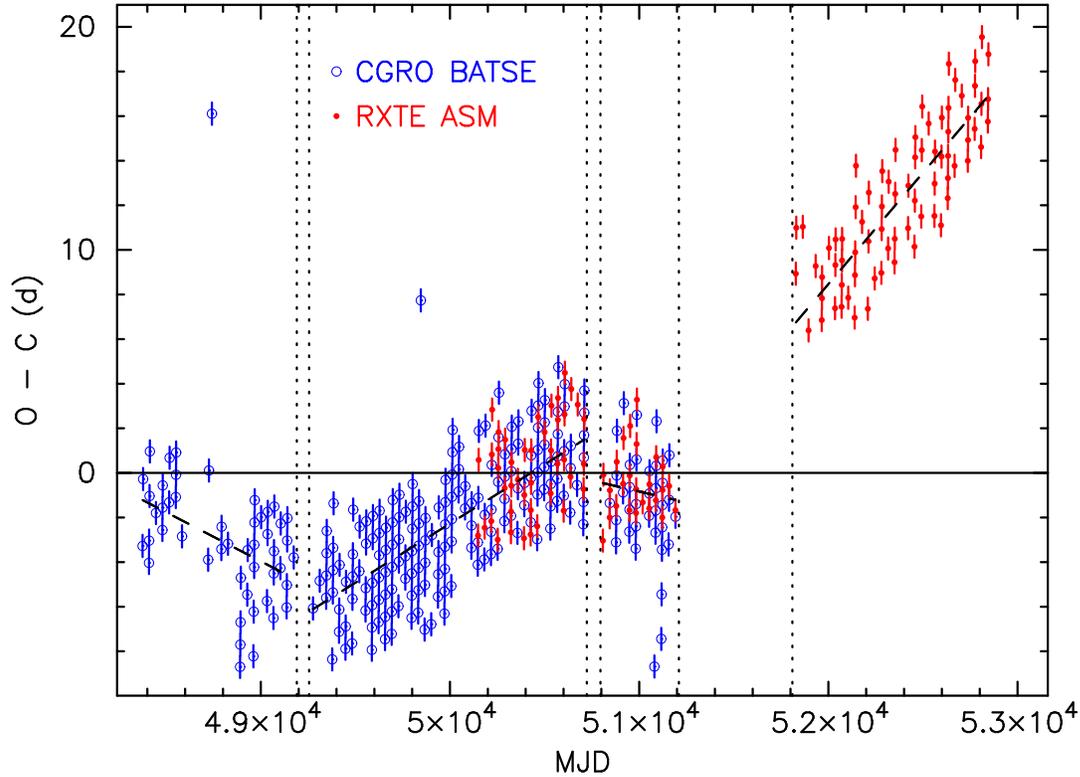}}
\noindent
\end{picture}
\vspace{100mm}
\figcaption{Observed $-$ Computed time series of the main-on states
in the  \her\ precession cycle.   Each point is  a daily  dwell in the
time series  that contains count rates  above a threshold of 0.05 \cs\
(BATSE)   or  5 \cs\    (ASM).   The reference    ephemeris adopted is
$T_{\mathrm  35} =  50986.7  +  34.79\,E_{\mathrm 35}$.  Dashed  lines
represent linear least-square fits to the $O-C$ residuals between each
ALS. Dotted vertical lines represent the  start and stop times of each
ALS.\label{fig:o-c}}
\end{figure*}

\begin{table*}
\begin{center}
\caption{Precession periods, \pprec, period derivative, \pprecdot, and 
maximum averaged main-on state count rates, \fmax, from four epochs of
the   \her\  precession   cycle.  Cycle  number   is  relative  to MJD
50\,986.7,        for           consistency          with        Fig.\
\ref{fig:o-c}. \vspace{5mm}\label{tab:epochs}}
{\footnotesize
\begin{tabular}{crrcccll}
\tableline\tableline
\multicolumn{1}{c}{Epoch} & \multicolumn{2}{c}{Cycle} & \multicolumn{2}{c}{MJD} & \multicolumn{1}{c}{\pprec} & \multicolumn{1}{c}{\pprecdot} & \multicolumn{1}{c}{\fmax} \\
& \multicolumn{1}{c}{Start} & \multicolumn{1}{c}{Stop} & \multicolumn{1}{c}{Start} & \multicolumn{1}{c}{Stop} & \multicolumn{1}{c}{(d)} & \multicolumn{1}{c}{(d\,d$^{-1}$)} & \multicolumn{1}{c}{(count\,s$^{-1}$)} \\
\tableline
1 & $-75$ & $-52$ & 48\,377 & 49\,213 & $34.637 \pm 0.010$ & $-(3 \pm 1)$\dex{-7} & $0.046 \pm 0.002$\tablenotemark{a} \\
2 & $-49$ &  $-8$ & 49\,282 & 50\,708 & $34.977 \pm 0.003$ & \hspace{.8em}$(3 \pm 2)\dex{-8}$ & $0.064 \pm 0.001$\tablenotemark{a}, $4.79 \pm 0.05$\tablenotemark{b} \\
3 & $-5$ &  $+6$ & 50\,813 & 51\,195 & $34.745 \pm 0.021$ & $-(7 \pm 7)\dex{-10}$ & $0.042 \pm 0.003$\tablenotemark{a}, $3.66 \pm 0.07$\tablenotemark{b} \\
4 & $+26$ &  $+54$ & 51\,892 & 52\,866 & $35.137 \pm 0.007$ & \hspace{.8em}$(1 \pm 1)\dex{-7}$ & \hspace{7.1em}$5.00 \pm  0.06$\tablenotemark{b} \\
\tableline
\end{tabular}
\tablenotetext{a}{\cgro\ BATSE}
\tablenotetext{b}{\xte\ ASM}}
\end{center}
\end{table*}

The extended ALS  also reveals itself in  the  hardness data.  A  soft
spectrum in the  ALS is consistent  with  that seen during normal  low
states of the 35d  cycle \citep{sco99,cla03}.  There is some  evidence
that  the hardness ratio returns to  a normal level before the main-on
cycle returns in count rate, and of  long-term hardness evolution over
the duration of ASM monitoring.   For an analysis  of the structure of
ASM  count  rates, folded   over  the 35d  cycle,  see  \citet{sha98},
\citet{sco99} and \citet{lea02}.    A  more detailed  analysis of  35d
spectral evolution with {\it GINGA} is presented by \citet{lea01}.

A period search over the ASM data obtained  after the 1999 ALS between
MJD  51\,892--52\,866  using  the Lomb-Scargle  method   \citep{sca82}
reveals a   best  period of $35.10   \pm 0.01$d,  with  a  false-alarm
probability  $<10^{-6}$.   This  compares directly   with an identical
search over the pre-1999 ASM data, MJD 50\,813--51\,195, of $34.79 \pm
0.03$d.  Clearly  there   is dichotomy   here.  To  qualify   this, we
construct  an ``Observed  $-$ Computed''  ($O-C$)  curve of precession
cycles (Fig.\ \ref{fig:o-c}).    Since turn-on times  are not strictly
periodic, the  simplest  reference point   for  this analysis  is  the
mid-point of each main-on state, $T_{\mathrm {main}}$.  Any data point
$>$ an arbitrary threshold of 0.05 (BATSE) and 5 (ASM) \cs\ is plotted
on the  $O-C$ diagram.  So each individual  main-on is  represented by
several data points in the diagram.  Scatter  is a relative measure of
the duration of main-on states, although ASM epochs cannot be compared
directly with BATSE epochs  in this way  because of the different flux
thresholds.  Our  reference  ephemeris,  measured  from all  ASM  data
before    the 1999 ALS,    is  $T_{\mathrm  {main}}$=  MJD $50986.7  +
34.79\,\eprec$, where \eprec\ is an  integer.  The quantity subtracted
from BATSE  and ASM  times to produce  the  $O-C$ time series  is  the
nearest $T_{\mathrm {main}}$ to each daily dwell.

There are at least four epochs  of different precession period evident
in Fig 2.  There is no evidence for episodes of high \pprecdot, except
at discrete times that  correspond to each  ALS.  The one exception is
at MJD 50\,750.   There have been no  reports of an ALS previously for
this date,  however  we suggest that the  low  count rates and  period
change at this epoch qualify it as such.  There  is an unfortunate gap
in the published \her\ pulse  period history \citep{par99} during this
epoch where one would expect an  episode of spin-down from the neutron
star.  Individual   Lomb-Scargle searches and   quadratic  fits to the
$O-C$  residuals  provide  independent,   and consistent, measures  of
\pprec\ for each epoch.    $O-C$ fits yield the   smallest statistical
error and we provide  these in Table \ref{tab:epochs}. Measured values
of \pprecdot\ are small during all epochs  and these are also included
in Table \ref{tab:epochs}.  Best fit  ephemerides are plotted over the
$O-C$ diagram as dashed  lines, although we have   made no attempt  to
align  the ephemerides at the  turning points.  Based on our arbitrary
count rate threshold, there are other candidates for ALS, particularly
during the  early  BATSE era  at MJD 48\,600.   However,  there  is no
evidence within the $O-C$    residuals or pulse history diagrams    to
confirm this.

\begin{figure*}
\begin{picture}(100,0)(10,20) 
\put(0,0){\includegraphics{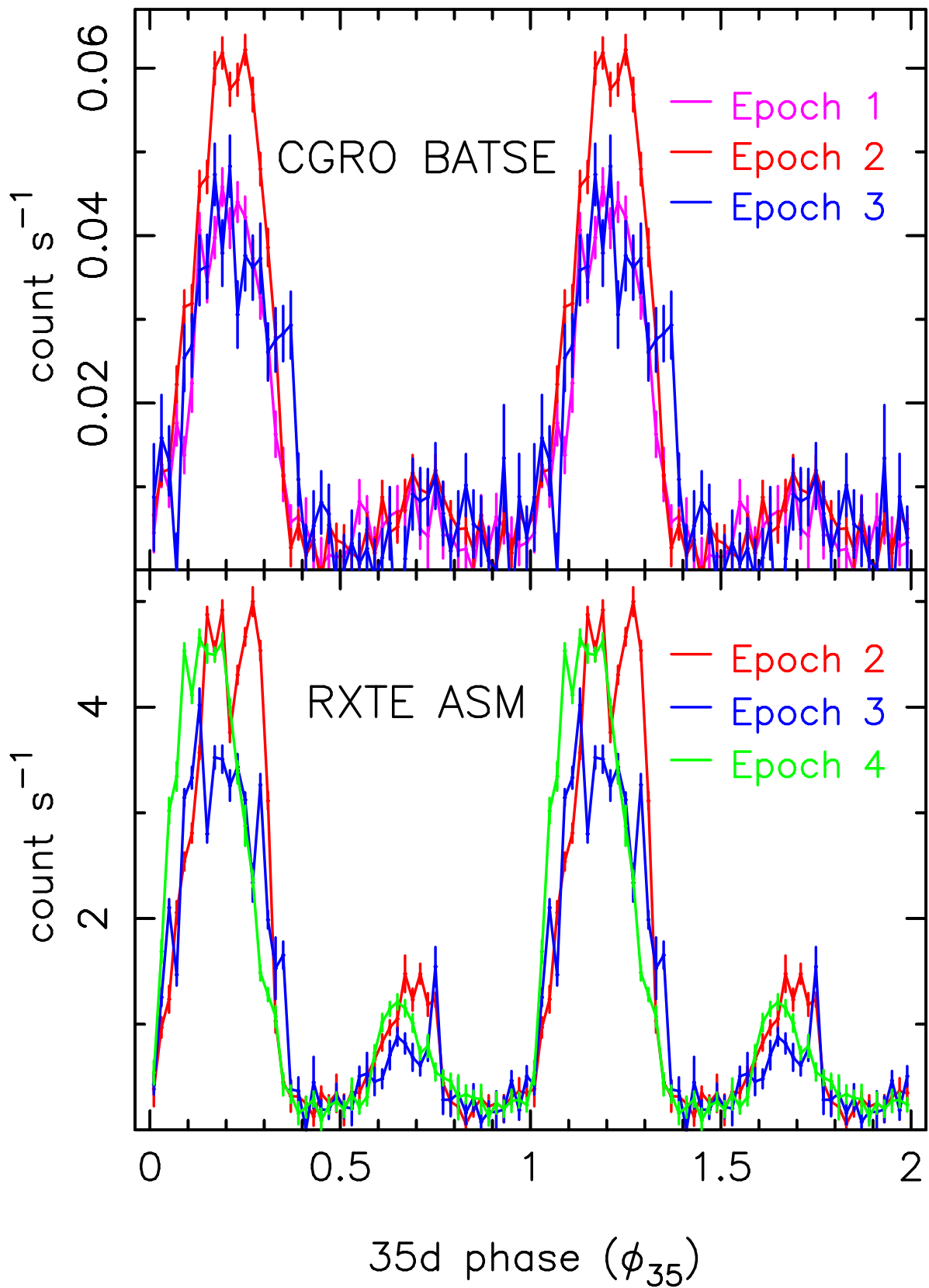}}
\noindent
\end{picture}
\vspace{200mm}
\figcaption{One-day BATSE and ASM dwells from each of the four epochs, 
folded over the epoch-dependent precession period and averaged into 50
phase bins.  The epoch  ranges are defined in Table  \ref{tab:epochs}.
The 35d cycle is plotted twice for the interest of clarity.
\label{fig:fold}}
\end{figure*}

\begin{figure*}
\begin{picture}(100,0)(10,20) 
\put(0,0){\includegraphics{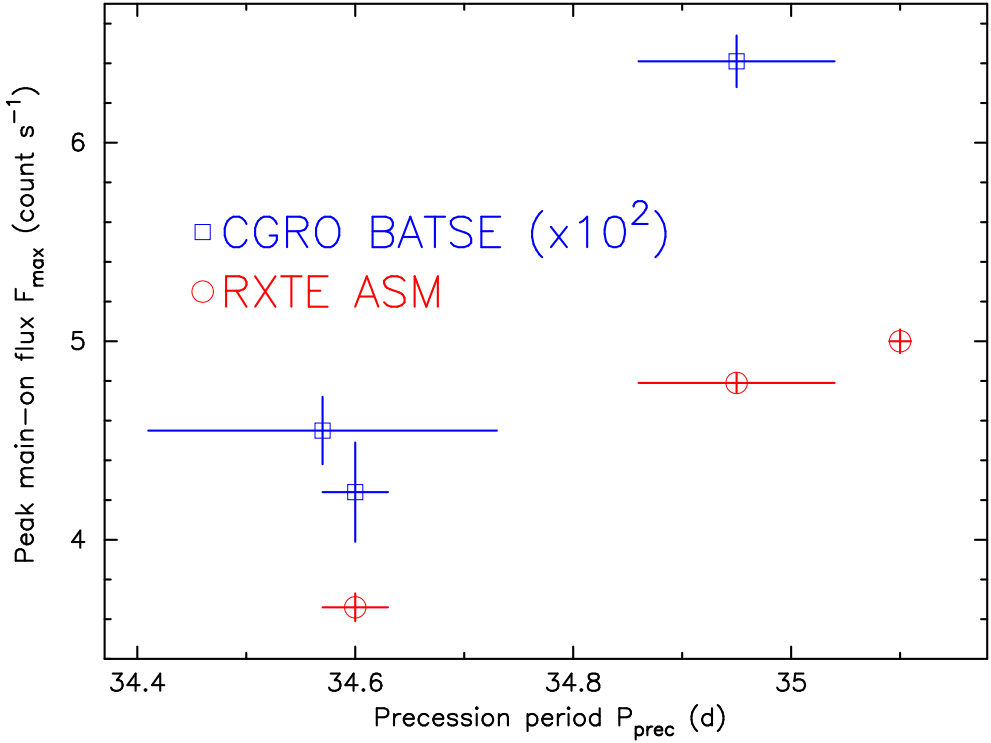}}
\noindent
\end{picture}
\vspace{81mm}
\figcaption{The time-averaged main-on BATSE and ASM count rates 
between  \phip\   = 0.1--0.4   as  a  function of  precession  period,
\pprec.  Since the two instruments  are  sensitive to different energy
ranges we have not renormalized one to  the other, but have multiplied
the  BATSE  count rates by   $10^2$  to incorporate  them  on the same
plot. No attempt is made to correlate BATSE fluxes with ASM fluxes due
to the discrepent energy ranges and instrument characteristics between
the two detectors.
\label{fig:correlation}}
\end{figure*}

Epoch-folding of the data over  \pprec, shows that turn-ons  typically
occur $0.4\pi$ in phase before the mid-point of main-on, as defined by
our threshold criteria, although there is an uncertainly of a few days
because the turn-ons  and maxima are  decoupled.  As discussed in Sec.
\ref{sec:introduction},   turn-ons appear  not    to  be  defined   by
precession period alone,   also  being  dependent on   orbital  phase.
Therefore, while  we  provide  a  precession ephemeris below   for the
turn-on  times   of  epoch 4,   $T_{\mathrm{on}}$, we    must  allow a
systematic uncertainty of $\sim$ 1 orbital cycle ($\sim$ 2d):
\begin{equation}
T_{\mathrm{on}} = \mbox{MJD 51\,821(2) + 35.10(1)\eprec}
\end{equation} 
Fig.\  \ref{fig:fold} plots  the  time series, separated   by epoch as
defined  in Table   \ref{tab:epochs},  and folded  over the   relevant
precession period.  The data  are averaged  into  50 phase bins, where
\phip\ is the precession  phase. \phip\ =  0 is defined as the turn-on
time.  There  are variations  in the main-on  structure from  epoch to
epoch, most noticeably in the peak flux of the main on state.

In Fig.\ \ref{fig:correlation}  we plot  main-on  count rates,  \fmax,
versus precession period, \pprec.  \fmax\ is measured as the peak of a
Gaussian  fit  of the  folded  light curves  in   Fig.\ \ref{fig:fold}
between \phip\  = 0.0--0.4.  We find  a positive  correlation  between
\pprec\ and \fmax\ in both the BATSE and ASM data, although the sample
is, of course, very small,  and the different  energy bands of the two
instruments means that they cannot be compared directly.

\section{Discussion}
\label{sec:discussion}

\citet{cla03} noticed the period variation in the \her\ precession 
cycle at MJD 50\,750, but interpreted it as  a phase shift, suggesting
that this event resulted eventually in the 1999 ALS, 400d later.  From
a longer  baseline of  data, we  offer the  alternative interpretation
that each ALS is accompanied by  a ``simultaneous'', change in \pprec.
The timescale  of state changes must  be just a few precession cycles.
These  are followed by epochs  where  \pprecdot $\simeq$ 0, indicating
long-lived, stable disk structure.   Extrapolating the quadratic $O-C$
fit for epoch 4, we  find that an  instantaneous \pprecdot\ would have
occurred  towards  the beginning of the  1999  ALS.  A Gabor transform
analysis (i.e.\ a short-time  Fourier transform with a sliding window;
\cite{hei89},  appropriate for uncovering  evolution in  frequency and
amplitude, resolves the  ASM light curve of Her  X-1 into a series  of
distinct episodes of constant period, consistent with the measurements
reported in  Table  \ref{tab:epochs}.   However,  due  to  the smaller
sampling, the period uncertainties resulting  from the Gabor transform
analysis   are much     larger     than  those reported    in    Table
\ref{tab:epochs}.  This  explains why the  three ASM precession epochs
were   not   distinguishable  in the   dynamical   power   analysis of
\citep{cla03}.

From Fig.\ \ref{fig:correlation},   we can speculate  that the main-on
flux and precession  period are related.  At the  peak of the  main-on
state, the absorbing column in front of the  pulsar is consistent with
the  galactic column \citep{fiu98}, which does  not  contribute to the
BATSE or ASM  energy  bands.  It  is reasonable, therefore,  to assume
that \fmax\ measures the accretion flux from the pulsar \citep{sco00}.
A relationship between \pprec\ and \fmax\ is a natural prediction from
models  of radiation-  and wind-driven precession \citep{sch94,wij99},
however an attempt to  correlate the two quantitatively  would require
radiative   and    hydrodynamic   models  of   enormous    complexity.
\citet{wij99} show that the precession  timescale ($P_{\mathrm d}$) is
governed by:
\begin{equation}
P_{\mathrm d} \sim 
\frac{\Sigma_{\mathrm d} R_{\mathrm d}^{3/2}}{L_{\mathrm d}}
\label{eq:timescale}
\end{equation}
where $R_{\mathrm d}$ is   a  characteristic accretion  disk   radius,
$\Sigma_{\mathrm d}$ a  surface   density and $L_{\mathrm  d}$  is the
strength of the radiation field from the disk.  If the radiation field
from the disk is  dominated by irradiation from  a central source then
\fmax\ $\propto$ $L_{\mathrm d}$.     Since we find  \fmax\  $\propto$
\pprec, Eqn.\ \ref{eq:timescale}  is not  consistent with observation.
However  disk warp  is  self-limiting;  brighter  than  some threshold
$L_{\mathrm  {d,max}}$, the warp becomes so tightly wound
that the disk is self-occulting and radiation pressure is reduced.  In
this phase space the  precession  period will be   directly-correlated
with the neutron star  flux, because  the surface area on
the disk incident to  the pulsar decreases  with increasing warp.  The
simplest picture for \her, therefore, is that the pulsar luminosity is
close to the stability limit.  This is consistent with the predictions
from analytical modeling by \citet{ogi01}.

A corollary from Eqn.  \ref{eq:timescale} is that it is of no surprise
that the precession period varies from epoch-to-epoch.  In fact, it is
predicted directly  from   theory \citep{wij99,ogi01}.    If structure
extends  to the  outer  disk edge, any  alteration  in disk  tilt will
result   in the deposition   of  mass by the    ballistic stream at  a
different  range of  accretion  disk radii  \citep{lub89,sch96}.   The
resulting redistribution of surface density over the disk will yield a
new  equilibrium  configuration for the  new epoch,  according to Eqn.
\ref{eq:timescale}.

We have determined \pprec\ from the entire time series rather than the
interval between  consecutive main-on turn-ons.   Since turn-ons occur
at only orbital  phases \phio $\simeq$ 0.2  and 0.7, it  is natural to
assume that structure locked in the rotational  frame of the binary is
partly responsible    for  turn-on  times   \citep{sch96}.   Measuring
intervals between consecutive turn-ons  would provide the same results
as Sec.\   \ref{sec:results} only if  a  large number  of  cycles were
sampled.  As we  have shown, \pprec\  has not remained stable for more
than   40 consecutive cycles in the   BATSE and ASM era.  Consequently
measuring   turn-on intervals  provides   a more uncertain measure  of
\pprec.  One   interesting direction  for  future   work would  be  to
determine  whether turn-on times  can be predicted  from \pprec, i.e.,
does the time of turn-on correspond to whichever orbital phase, 0.2 or
0.7, immediately follows an   undetermined reference phase within  the
stable  precession cycle?  Of course, the  reference will change after
each ALS.

\section{Conclusions}
\label{sec:conclusions}

We  have  redefined the accretion  disk  precession ephemeris  for the
X-ray pulsar \her\ between the end of the 1999--2000 ALS and the start
of the 2003--2004 ALS event.   Relative to the most recently published
work  on  cycle  timing  \citep{man03}, the  new ephemeris  provides a
correction to the  turn-on time of the  precession cycle of $\sim 14$d
at the time of  writing, critical  to  research teams  and observatory
schedulers during the  current  ALS. A period  analysis  of the \cgro\
BATSE  and \xte\  ASM time series   since 1991 reveals  four epochs of
stable precession period. The  precession epochs are each separated by
an ALS and the change of period appears to be instantaneous within the
resolution  of the  $O-C$  diagram,  i.e.   a  few precession  cycles.
Testable predictions are that  \her\  will return from  the 2003--2004
ALS with a new precession period and epoch-averaged peak main-on flux.

\acknowledgements
Results provided by the \xte\ ASM teams at MIT and at the RXTE SOF and
GOF at  NASA's GSFC,  and the  \cgro\ BATSE instrument  team at NASA's
MSFC.

\end{document}